\documentclass[10pt,twocolumn,amssymb,amsmath,aps,pra]{revtex4-1}
\usepackage{graphicx}

\newcommand*{\Scale}[2][4]{\scalebox{#1}{$#2$}}

\begin{document}

\title{Unified position-dependent photon-number quantization in layered structures}
\date{\today}
\author{Mikko Partanen}
\author{Teppo H\"ayrynen}
\author{Jani Oksanen}
\author{Jukka Tulkki}
\affiliation{Department of Biomedical Engineering and Computational Science, Aalto University, P.O. Box 12200, 00076 Aalto, Finland}

\begin{abstract}
We have recently developed a position-dependent quantization scheme for
describing the ladder and effective photon-number operators associated with the electric field
to analyze quantum optical energy transfer in lossy and 
dispersive dielectrics [Phys. Rev. A, 89, 033831 (2014)].
While having a simple connection to the thermal balance of the system, these operators
only described the electric field and its coupling to lossy dielectric bodies.
Here we extend this field quantization scheme to include the magnetic field and
thus to enable description of the total electromagnetic field
and discuss conceptual measurement schemes to verify the predictions.
In addition to conveniently describing the formation of thermal balance, the generalized
approach allows modeling of the electromagnetic pressure and Casimir forces.
We apply the formalism to study the local steady state field temperature
distributions and electromagnetic force density in cavities with cavity
walls at different temperatures. The calculated local electric and magnetic
field temperatures exhibit oscillations that depend on the position as well
as the photon energy. However, the effective photon number and field temperature
associated with the total electromagnetic field is always position-independent in
lossless media. Furthermore, we show that the direction of the electromagnetic force
varies as a function of frequency, position, and material thickness.
\end{abstract}

\maketitle

\section{Introduction}
The quantum optical processes and field quantization in lossy structures exhibiting interference
represent a fascinating challenge to our understanding of wave-particle dualism, interwined
electric and magnetic fields, and wave-matter interactions. The quantization of the electromagnetic (EM)
field in a dielectric medium has been widely studied during the last few decades especially in
layered structures \cite{Knoll1991,Allen1992,Huttner1992,Barnett1995,Matloob1995,Matloob1996}.
It has been established that the field operators obey the well-known canonical commutation relations
\cite{Barnett1995,Matloob1995}, but there are reported anomalies in the commutation relations
of the intracavity ladder operators \cite{Raymer2013b,Barnett1996,Ueda1994,Aiello2000,Stefano2000}
leading to difficulties in defining a well-behaving photon number.
We have recently introduced photon ladder operators associated with the electric field
in a way that is consistent with the canonical commutation relations
and gives further insight on the local effective photon number, thermal balance, and the
formation of the local thermal equilibrium \cite{Partanen2014a,Partanen2014b}.
This approach, however, neglects the magnetic contributions that are important
in determining the properties of the total EM field, its energy density,
and EM pressure.

In this paper, we extend our ladder operator formalism to
describe also the magnetic field
and the total EM field in layered structures.
We present the photon ladder and number operators for the electric and magnetic fields
and express the photon number of the total EM field in terms of the electric
and magnetic photon-number operators. The electric and
magnetic field associated effective photon numbers generally oscillate even in the vacuum, but we
show that the oscillations balance each other so that the photon number associated with the total EM field is
constant in the vacuum as expected. We also establish the relation of the local photon-number operators
to the field temperatures, EM pressures, and thermal balance of the system
providing insight on the physical interpretation of the quantities.
This is followed by applying the presented theoretical concepts to study the position and photon energy
dependence of the effective electric, magnetic, and total EM field temperatures
and corresponding local densities of states (LDOS) in a geometry of a vacuum cavity
formed between two semi-infinite thermal reservoir media at different temperatures.
In addition, the relation of the effective photon number to the EM pressure is studied
by examining the force exerted on lossy and lossless material slabs placed inside the vacuum cavity.

\vspace{-0.6cm}
\section{\label{sec:theory}Field quantization}

\vspace{-0.3cm}
\subsection{Noise operator formalism}
\vspace{-0.4cm}

The theoretical foundations of the present work originate from the noise operator
formalism developed by Matloob \emph{et al.} \cite{Matloob1995,Matloob1996}.
In the noise operator formalism, the field operators are expressed in terms of the
Green's function $G(x,\omega,x')$ of the Helmholtz equation and the bosonic source
field operator $\hat f(x,\omega)$ obeying $[\hat f(x,\omega),\hat f^\dag(x',\omega')] = \delta(x-x')\delta(\omega-\omega')$
describing the material state.
For example, the equation for the positive frequency part of the vector potential operator is given by
$\hat{A}^+(x,\omega) = \int_{-\infty}^\infty G_\mathrm{A}(x,\omega,x')\hat f(x',\omega)dx'$,
where $G_\mathrm{A}(x,\omega,x')=\mu_0j_0(x',\omega) G(x,\omega,x')$, in which
$j_0(x,\omega)=\sqrt{4\pi\hbar\omega^2\varepsilon_0\mathrm{Im}[n(x,\omega)^2]/S}$
is a scaling factor, $n(x,\omega)$ is the refractive index of the medium,
$\hbar$ is the reduced Planck's constant, $\varepsilon_0$ is the permittivity
of vacuum, $\mu_0$ is the permeability
of vacuum, and $S$ is the area of quantization in the $y$-$z$ plane \cite{Matloob1995}.
Similar expressions are valid for electric and magnetic field operators $\hat{E}^+(x,\omega)$
and $\hat{B}^+(x,\omega)$ with $G_\mathrm{A}(x,\omega,x')$ replaced by
$G_\mathrm{E}(x,\omega,x')=i\omega\mu_0 j_0(x',\omega)G(x,\omega,x')$ and
$G_\mathrm{B}(x,\omega,x')=\mu_0 j_0(x',\omega)\partial G(x,\omega,x')/\partial x$
as defined in Ref.~\cite{Partanen2014a}.

\subsection{\label{sec:operators}Photon operators}
\vspace{-0.4cm}

In any quantum electrodynamics (QED) description, the canonical commutation relations are satisfied for
field quantities, i.e., $[\hat A(x,t),\hat E(x',t)]=-i\hbar/(\varepsilon_0S)\delta(x-x')$ \cite{Scheel1998},
but the same is not generally true for the canonical commutation relations of the ladder operators.
The dominant approach in evaluating the ladder operators has been to separate the
field operators obtained from QED either into the left and right 
propagating normal modes or into the normal modes related to the left and right inputs
and the corresponding ladder operators \cite{Barnett1996,Gruner1996,Aiello2000}.
This is tempting in view of the analogy with classical EM, but in most cases results in
ladder operators that are not unambiguously determined due to the possibility to scale
the normal modes nearly arbitrarily. Also other definitions based on separately accounting for
the noise contribution have been reported \cite{Stefano2000}, but they do not result in the
canonical commutation relations for the ladder operators either.

We have very recently presented a quantization scheme that uses the requirement of preservation of the
canonical commutation relation $[\hat a(x,\omega),\hat a^\dag(x,\omega')]=\delta(\omega-\omega')$
as a starting point for defining the photon annihilation operator that describes the electric
field \cite{Partanen2014a,Partanen2014b}.
In the developed formalism the electric field annihilation operator is given by \cite{Partanen2014b}
\begin{align}
 \hat a_\mathrm{e}(x,\omega) &=\sqrt{\frac{\varepsilon_0}{2\pi^2\hbar\omega\rho_\mathrm{e}(x,\omega)}}\hat E^+(x,\omega)\nonumber\\
 &= \sqrt{\frac{\varepsilon_0}{2\pi^2\hbar\omega\rho_\mathrm{e}(x,\omega)}}\int_{-\infty}^\infty G_\mathrm{E}(x,\omega,x')\hat f(x',\omega)dx'
 \label{eq:electrica},
\end{align}
where the factor $\rho_\mathrm{e}(x,\omega)$ has been shown to correspond to the 
conventional definition of the electric contribution of the
local density of EM states (electric LDOS) defined as \cite{Joulain2003}
\begin{align}
 \rho_\mathrm{e}(x,\omega) &= \frac{\varepsilon_0}{2\pi^2\hbar\omega}\int_{-\infty}^\infty|G_\mathrm{E}(x,\omega,x')|^2dx'\nonumber\\
 &=\frac{2\omega^3}{\pi c^4S}\int_{-\infty}^\infty\mathrm{Im}[n(x',\omega)^2]|G(x,\omega,x')|^2dx'\nonumber\\
 &=\frac{2\omega}{\pi c^2S}\mathrm{Im}[G(x,\omega,x)]
 \label{eq:eldos},
\end{align}
where $c$ is the speed of light in vacuum.

In this work, we unify the previously introduced electric-field-based photon-number
picture to the description of the total EM field. For this purpose,
we first define a magnetic field annihilation operator proportional to the magnetic field operator
written for positive frequencies as $\hat B^+(x,\omega)=C(x,\omega)\hat a_\mathrm{m}(x,\omega)$.
Just like in the case of the electric field \cite{Partanen2014a},
the normalization coefficient $C(x,\omega)$ is defined by using the requirement that the canonical
commutation relation is fulfilled leading to the relation,
\begin{align}
 \hat a_\mathrm{m}(x,\omega) &=\sqrt{\frac{\varepsilon_0c^2}{2\pi^2\hbar\omega\rho_\mathrm{m}(x,\omega)}}\hat B^+(x,\omega)\nonumber\\
  &= \sqrt{\frac{\varepsilon_0c^2}{2\pi^2\hbar\omega\rho_\mathrm{m}(x,\omega)}}\int_{-\infty}^\infty\!\! G_\mathrm{B}(x,\omega,x')\hat f(x',\omega)dx'
 \label{eq:magnetica}.
\end{align}
In Eq.~\eqref{eq:magnetica}, the magnetic contribution of the
local density of EM states (magnetic LDOS) is given by
\begin{align}
 \rho_\mathrm{m}(x,\omega) &= \frac{\varepsilon_0 c^2}{2\pi^2\hbar\omega}\int_{-\infty}^\infty|G_\mathrm{B}(x,\omega,x')|^2dx'\nonumber\\
 &=\frac{2\omega^3}{\pi c^4S}\int_{-\infty}^\infty\mathrm{Im}[n(x',\omega)^2]\Big|\frac{\partial G(x,\omega,x')}{k_0\partial x}\Big|^2dx'\nonumber\\
 &=\frac{2\omega}{\pi c^2S}\mathrm{Im}[n(x,\omega)^2G_{\{-r\}}(x,\omega,x)]
 \label{eq:bldos},
\end{align}
where $k_0=\omega/c$ and $G_{\{-r\}}(x,\omega,x)$
has been defined as an auxiliary Green's function calculated for a structure
where all reflection coefficients have been transformed by $r(\omega)\longrightarrow -r(\omega)$.
The transformation enables writing the magnetic LDOS expression in a form resembling the electric LDOS.
In homogeneous media, the electric and magnetic field annihilation operators are
identical, but generally the annihilation operator for the magnetic
field is different from the annihilation operator for the electric
field due to different spatial dependence of electric and magnetic fields
as seen from the definitions of $G_\mathrm{E}(x,\omega,x')$ and $G_\mathrm{B}(x,\omega,x')$.

The electric and magnetic photon-number operators are obtained from the ladder operators as
$\hat n_i(x,\omega)=\int\hat a_i^\dag(x,\omega)\hat a_i(x,\omega')d\omega'$, $i\in\{\mathrm{e},\mathrm{m}\}$,
and, therefore, their expectation values are given in terms of the Green's functions and the source field photon-number
operator as
\begin{align}
\Scale[0.96]{\displaystyle\langle\hat n_\mathrm{e}(x,\omega)\rangle} & =\Scale[0.96]{\displaystyle\frac{\varepsilon_0}{2\pi^2\hbar\omega\rho_\mathrm{e}(x,\omega)}\!\int_{-\infty}^\infty\! |G_\mathrm{E}(x,\omega,x')|^2\langle\hat\eta(x',\omega)\rangle dx' }\nonumber\\
 &=\Scale[0.96]{\displaystyle\frac{2\omega^3}{\pi c^4S\rho_\mathrm{e}(x,\omega)}\!\int_{-\infty}^\infty\!\mathrm{Im}[n(x',\omega)^2]|G(x,\omega,x')|^2}\nonumber\\
 &\hspace{0.4cm}\Scale[0.96]{\displaystyle\times\langle\hat\eta(x',\omega)\rangle dx'},
 \label{eq:electricn}
\end{align}
\begin{align}
\Scale[0.95]{\displaystyle\langle\hat n_\mathrm{m}(x,\omega)\rangle} & =\Scale[0.95]{\displaystyle\frac{\varepsilon_0c^2}{2\pi^2\hbar\omega\rho_\mathrm{m}(x,\omega)}\!\int_{-\infty}^\infty\! |G_\mathrm{B}(x,\omega,x')|^2\langle\hat\eta(x',\omega)\rangle dx'}\nonumber\\
 &=\Scale[0.95]{\displaystyle\frac{2\omega^3}{\pi c^4S\rho_\mathrm{m}(x,\omega)}\!\int_{-\infty}^\infty\!\mathrm{Im}[n(x',\omega)^2]\Big|\frac{\partial G(x,\omega,x')}{k_0\partial x}\Big|^2}\nonumber\\
 &\hspace{0.4cm}\Scale[0.95]{\displaystyle\times\langle\hat\eta(x',\omega)\rangle dx'},
 \label{eq:magneticn}
\end{align}
where we have defined the source field photon-number operator as
$\hat \eta(x,\omega) = \int\hat f^\dag(x,\omega)\hat f(x',\omega')\,dx'd\omega'$
and assumed that the noise operators at different positions and at different frequencies
are uncorrelated. For media in local thermal equilibrium, the source field photon-number expectation value
at position $x$ is
\begin{equation}
 \langle\hat \eta(x,\omega)\rangle = \frac{1}{e^{\hbar\omega/(k_\mathrm{B}T(x))}-1},
 \label{eq:sourcefieldn}
\end{equation}
where $k_\mathrm{B}$ is the Boltzmann constant and
$T(x)$ is the temperature distribution of the medium.

The expectation value of the total photon-number operator describing the energy quanta of the
total EM field is obtained either by using a similar quantization scheme as for electric and magnetic fields
or directly written as an LDOS weighted sum of the electric and magnetic photon numbers as
\begin{align}
 &\langle\hat n_\mathrm{tot}(x,\omega)\rangle\nonumber\\
 &=\frac{|n(x,\omega)|^2\rho_\mathrm{e}(x,\omega)\langle\hat n_\mathrm{e}(x,\omega)\rangle+\rho_\mathrm{m}(x,\omega)\langle\hat n_\mathrm{m}(x,\omega)\rangle}{|n(x,\omega)|^2\rho_\mathrm{e}(x,\omega)+\rho_\mathrm{m}(x,\omega)}\nonumber\\
 &=\frac{\omega^3|n(x,\omega)|^2}{\pi c^4S\rho_\mathrm{tot}(x,\omega)}\int_{-\infty}^\infty\mathrm{Im}[n(x',\omega)^2]\Big(|G(x,\omega,x')|^2\nonumber\\
 &\hspace{0.4cm}+\Big|\frac{\partial G(x,\omega,x')}{k(x,\omega)\partial x}\Big|^2\Big)\langle\hat\eta(x',\omega)\rangle dx',
 \label{eq:totaln}
\end{align}
where $k(x,\omega)=\omega n(x,\omega)/c$ and the total EM LDOS is the sum of the electric and magnetic contributions in
Eqs.~\eqref{eq:eldos} and \eqref{eq:bldos} given by
\begin{align}
 \rho_\mathrm{tot}(x,\omega) &=\frac{\omega^3|n(x,\omega)|^2}{\pi c^4S}\int_{-\infty}^\infty\mathrm{Im}[n(x',\omega)^2]\nonumber\\
 &\hspace{0.4cm}\times\Big(|G(x,\omega,x')|^2+\Big|\frac{\partial G(x,\omega,x')}{k(x,\omega)\partial x}\Big|^2\Big)dx'.
 \label{eq:uldos}
\end{align}
The total field photon number in Eq.~\eqref{eq:totaln} will be shown to be related to the total field energy and EM pressure
providing a meaningful definition for the photon number of the total EM field.
In contrast to the corresponding electric quantities,
the above total EM LDOS and the photon-number expectation value
are always constant in lossless media, as can be easily shown by calculating the
derivatives $\partial\rho_\mathrm{tot}(x,\omega)/\partial x$ and
$\partial\langle\hat n_\mathrm{tot}(x,\omega)\rangle/\partial x$,
which are equal to zero when the nonhomogeneous Helmholtz equation is fulfilled and
the refractive index at position $x$ is $n(x,\omega)=1$.
In the case of thermal fields, the photon-number operators in Eqs.~\eqref{eq:electricn},
\eqref{eq:magneticn}, and \eqref{eq:totaln} also allow one to calculate effective
local electric, magnetic, and total field temperatures as
\begin{equation}
 T_i(x,\omega)=\frac{\hbar\omega}{k_\mathrm{B}\ln[1+1/\langle\hat n_i(x,\omega)\rangle]}, \hspace{0.5cm}i\in\{\mathrm{e},\mathrm{m},\mathrm{tot}\}.
 \label{eq:temperature}
\end{equation}

\vspace{-0.4cm}
\subsection{\label{sec:balance}Thermal balance}
\vspace{-0.4cm}

A particularly insightful view of the electric field photon number is provided by its connection
to local thermal balance between the field and matter \cite{Partanen2014a}.
The spectral net emission rate can be compactly written as a product of the electric LDOS
and the difference of the local source field photon number and the electric field photon number as \cite{Partanen2014a}
\begin{equation}
\Scale[0.95]{\displaystyle\langle Q(x,t)\rangle_\omega=\hbar\omega^2\mathrm{Im}[n(x,\omega)^2]\rho_\mathrm{e}(x,\omega)[\langle\hat\eta(x,\omega)\rangle-\langle\hat n_\mathrm{e}(x,\omega)\rangle]}.
 \label{eq:divP}
\end{equation}
In resonant systems where the energy exchange is dominated by a narrow frequency band,
condition $\langle\hat Q(x,\omega)\rangle_\omega=0$ can be used
to  approximately determine the steady-state temperature of a weakly interacting resonant
particle \cite{Bohren1998}. This suggests that the electric field temperature is
experimentally measurable
by measuring the steady-state temperature reached by a detector with a weak
field-matter interaction that is dominated by the
coupling to the electric field.
Similar temperature measurement setup to measure the magnetic field photon number
is expected to be also possible using materials whose field-matter interactions have been engineered
to be sensitive to magnetic fields instead of electric fields, using,
e.g., micro-coil sensors \cite{Tumanski2007} or magnetic metamaterials
\cite{Pendry1999} at least at microwave frequencies.
Another possible measurement setup for the local electric field temperature
could, for example, use the transparent intracavity
photodetector studied by Lazar \emph{et al.} \cite{Lazar2011,Hola2013}.
Measuring the electric field at equilibrium conditions using a movable transparent
intracavity detector allows one to determine the electric LDOS since the photon number
is known and constant. Then changing the cavity wall temperatures and using the determined LDOS allows one
to calculate the position-dependent photon number from the measured field.
Measuring the magnetic field by using a similar scheme
is not straightforward at optical frequencies, but the existence of the
predicted phenomena could be demonstrated at microwave frequencies by using micro-coil sensors
that practically do not disturb the measured magnetic field \cite{Tumanski2007}.
The electric and magnetic field LDOSs and photon numbers together also determine
the total EM field quantities.

\vspace{-0.4cm}
\subsection{Energy density and EM pressure}
\vspace{-0.4cm}

Despite its natural connection to the energy balance, the photon-number operator
related to the electric field can exhibit oscillatory behavior, e.g.,
in vacuum cavities \cite{Partanen2014a,Partanen2014b}.
The total photon-number operator presented in Eq.~\eqref{eq:totaln} does not share
the same peculiarity as it accounts for both the electric and magnetic contributions,
which balance out the oscillations. To further study the physical significance of
the total field photon number, we will next discuss the energy density associated
with the electric and magnetic contributions, and their relation to the total
energy density and EM pressure.

The electric and magnetic field fluctuations and the total energy density
$\langle\hat u(x,t)\rangle_\omega=\varepsilon_0|n(x,\omega)|^2\langle\hat E(x,t)\rangle_\omega/2+\langle\hat B(x,t)\rangle_\omega/(2\mu_0)$
for a single polarization and angular frequency $\omega$ in terms
of the position-dependent photon-number operators are given by
\begin{align}
 \langle\hat E(x,t)^2\rangle_\omega & =\frac{\hbar\omega}{\varepsilon_0}\rho_\mathrm{e}(x,\omega)\Big(\langle\hat n_\mathrm{e}(x,\omega)\rangle+\frac{1}{2}\Big)\label{eq:efluct},\\[8pt]
 \langle\hat B(x,t)^2\rangle_\omega & =\frac{\hbar\omega}{\varepsilon_0c^2}\rho_\mathrm{m}(x,\omega)\Big(\langle\hat n_\mathrm{m}(x,\omega)\rangle+\frac{1}{2}\Big)\label{eq:bfluct},\\[8pt]
 \langle\hat u(x,t)\rangle_\omega & = \hbar\omega\rho_\mathrm{tot}(x,\omega)\Big(\langle\hat n_\mathrm{tot}(x,\omega)\rangle+\frac{1}{2}\Big)\label{eq:edensity}.
\end{align}
In defining the total energy density, we have used the definition accounting
for the energy of the polarizability of the matter. Furthermore, the media are assumed
to be nonmagnetic. The magnitudes of the field
fluctuations calculated from Eqs.~\eqref{eq:efluct} and \eqref{eq:bfluct} are
continuous at interfaces as required by the boundary conditions, but the energy density
in Eq.~\eqref{eq:edensity} can be discontinuous due to the effect of
material polarizability.

Classically, EM forces are calculated using Maxwell's stress tensor \cite{Jackson1999,Landau1984}.
We apply the Maxwell's stress tensor in the form,
\begin{equation}
 \overset{\text{\tiny$\leftrightarrow$}}{\mathbf{T}}(x,t)=\varepsilon_0|n(x,\omega)|^2\hat{E}(x,t)^2\mathbf{y}\mathbf{y}+\frac{1}{\mu_0}\hat{B}(x,t)^2\mathbf{z}\mathbf{z}-\hat u(x,t)\overset{\text{\tiny$\leftrightarrow$}}{\mathbf{I}},
 \label{eq:stress}
\end{equation}
where the classical fields have been replaced by their quantum analogs and
$\overset{\text{\tiny$\leftrightarrow$}}{\mathbf{I}}$ is the unit dyadic presented in the Cartesian
basis $(\mathbf{x},\mathbf{y},\mathbf{z})$ as
$\overset{\text{\tiny$\leftrightarrow$}}{\mathbf{I}}=\mathbf{x}\mathbf{x}+\mathbf{y}\mathbf{y}+\mathbf{z}\mathbf{z}$.
The mechanical force density operator is given by \cite{Novotny2006,Maggiore2005}
\begin{equation}
 \hat{\mathbf{F}}(x,t)=\nabla\cdot\overset{\text{\tiny$\leftrightarrow$}}{\mathbf{T}}(x,t)-\frac{1}{c^2}\frac{\partial}{\partial t}\hat{\mathbf{S}}(x,t),
 \label{eq:forcedensity}
\end{equation}
where $\hat{\mathbf{S}}(x,t)$ is the Poynting vector operator and the last term in Eq.~\eqref{eq:forcedensity}
gives the force density experienced by the EM field.
In the steady state the expectation value of the last term is zero since
the expectation value of the Poynting vector does not change in time.
In one dimension the $x$ component of the spectral force density expectation value then becomes
\begin{align}
 \langle\hat F_x(x,t)\rangle_\omega & =-\frac{\partial}{\partial x}\langle\hat u(x,t)\rangle_\omega\nonumber\\
 & =-\frac{\hbar\omega}{2}\Big(\frac{\partial}{\partial x}\rho_\mathrm{tot}(x,\omega)\Big)
  -\hbar\omega\Big(\frac{\partial}{\partial x}\rho_\mathrm{tot}(x,\omega)\Big)\nonumber\\
 &\hspace{0.4cm}\times\!\langle\hat n_\mathrm{tot}(x,\omega)\rangle-\hbar\omega\rho_\mathrm{tot}(x,\omega)\frac{\partial}{\partial x}\langle\hat n_\mathrm{tot}(x,\omega)\rangle.
 \label{eq:forcedensity2}
\end{align}
The first term corresponds to the familiar zero-point Casimir force (ZCF) \cite{Rodriguez2011,Antezza2008},
the second term is known as the thermal Casimir force (TCF) \cite{Passante2007,Sushkov2011,Klimchitskaya2008},
and the last term arising from the changes in the total photon number is called
a nonequilibrium Casimir force (NCF) since it disappears at thermal
equilibrium when the derivative of the photon number is zero.
The net force on any solid object extending from $x_1$ to $x_2$ can then be obtained
as $\langle\hat{\mathcal{F}}(t)\rangle_\omega/S=\int_{x_1}^{x_2}\langle\hat F_x(x,t)\rangle_\omega dx$.

The net force can be also obtained by using the concept of EM pressure.
The EM pressure along the $x$ direction is obtained directly from the stress
tensor as $-\overset{\text{\tiny$\leftrightarrow$}}{\mathbf{T}}_{xx}$, giving
\begin{equation}
 \langle\hat{p}(x,t)\rangle_\omega=\hbar\omega\rho_\mathrm{tot}(x,\omega)\Big(\langle\hat n_\mathrm{tot}(x,\omega)\rangle+\frac{1}{2}\Big).
 \label{eq:pressure}
\end{equation}
Therefore, the net force on an object extending from $x_1$ to $x_2$
can be obtained as $\langle\hat{\mathcal{F}}(t)\rangle_\omega/S=\langle\hat{p}(x_1,t)\rangle_\omega-\langle\hat{p}(x_2,t)\rangle_\omega$.

In small cavities the zero-point Casimir force usually dominates in the EM force.
A possible measurement setup in which the thermal and nonequilibrium contributions would be essential
is, for example, the measurement of force exerted on a material slab placed in the middle of a symmetric cavity
with boundaries at different temperatures. In this case, the densities of states
on the different boundaries outside the slab would be equal
[$\rho_\mathrm{tot}(x_1,\omega)=\rho_\mathrm{tot}(x_2,\omega)$] due to the symmetry, and therefore, the 
zero-point Casimir
contribution would cancel out. Then the spectral force on the slab simplifies to
\begin{equation}
 \frac{\langle\hat{\mathcal{F}}(t)\rangle_\omega}{S}=\hbar\omega\rho_\mathrm{tot}(x_1,\omega)\Big(\langle\hat n_\mathrm{tot}(x_1,\omega)\rangle-\langle\hat n_\mathrm{tot}(x_2,\omega)\rangle\Big).
 \label{eq:forceslab2}
\end{equation}
Measuring forces on cavity walls also provides a possible scheme for determining
the total EM LDOS and the total EM photon number inside the cavity.
Knowing the source field photon numbers and measuring
the forces (1) at equilibrium and (2) varying one of the reservoir temperatures,
one can unambiguously solve the unknown LDOSs and the EM photon number inside the
cavity. This is simplest in the nearly monochromatic case where the emissivity
of the reservoirs are concentrated over a narrow spectrum, but it should
also be possible in the more general case.
The total EM LDOS and the total photon number can also be determined
inside a cavity structure by separately measuring the electric and magnetic field
LDOSs and photon numbers, which together determine the total field quantities
as discussed in Sec.~\ref{sec:balance}.

\begin{figure*}
\includegraphics[width=\textwidth]{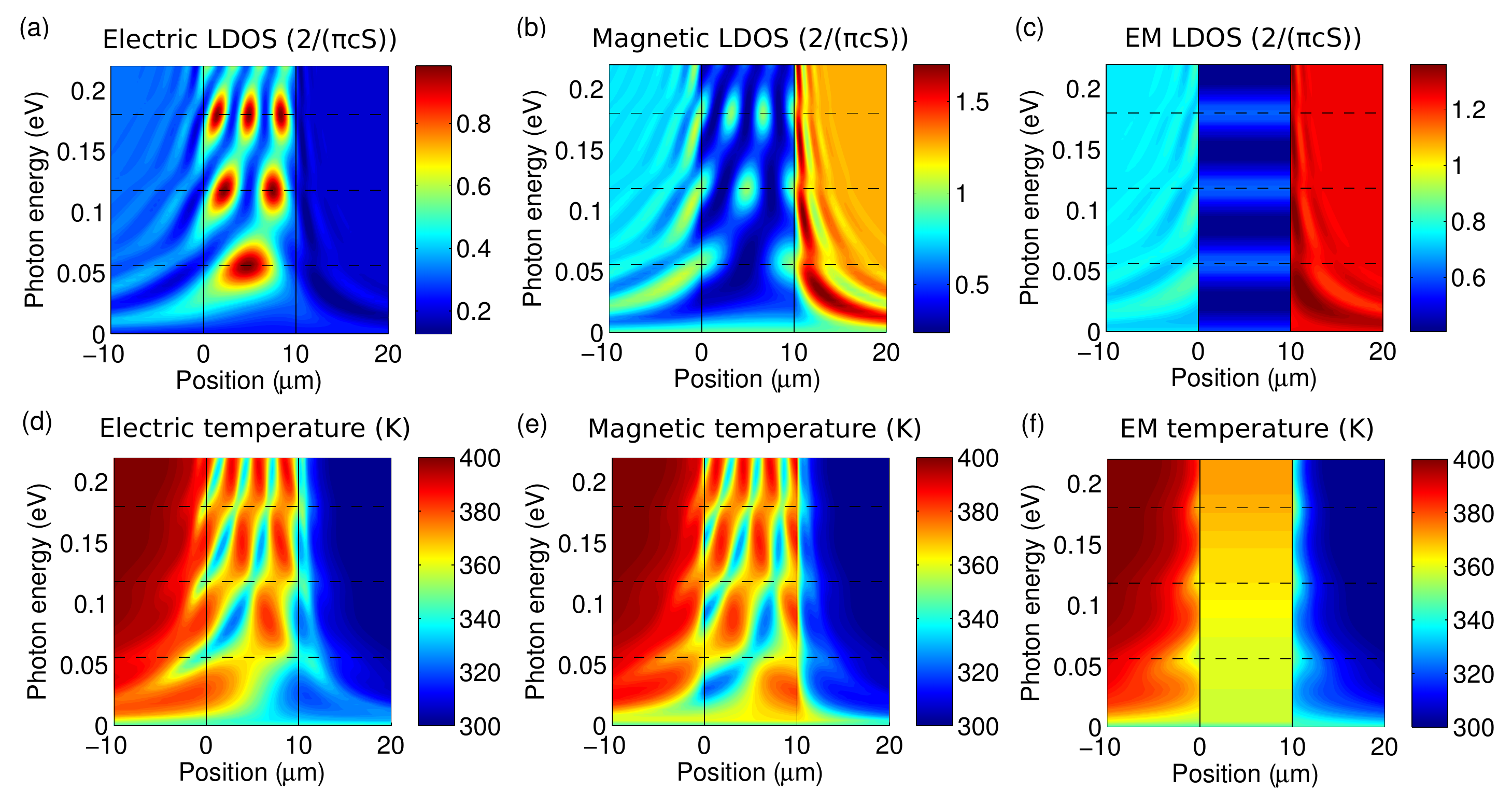}
\caption{\label{fig:temp}(Color online) (a) Electric LDOS, (b) magnetic LDOS, (c) total EM LDOS, (d)
electric field temperature, (e) magnetic field temperature, and (f) total field temperature
in the vicinity of a vacuum gap separating
lossy media with refractive indices $n_1=1.5+0.3i$ and $n_2=2.5+0.5i$ at temperatures 400 and 300 K.
Solid lines denote the boundaries of the cavity and dashed lines denote resonant energies.
The LDOSs are given in the units of $2/(\pi c S)$.}
\end{figure*}

\vspace{-0.6cm}
\section{\label{sec:results}Results}
\vspace{-0.3cm}

\subsection{Field temperatures}
\vspace{-0.4cm}

To investigate the physical implications of the concepts presented in Sec.~\ref{sec:theory}
we study the properties of the effective field temperatures and the corresponding local densities of states
in a vacuum cavity formed between two
semi-infinite media with refractive indices $n_1=1.5+0.3i$ and $n_2=2.5+0.5i$.
The temperatures of the left and right thermal reservoirs formed by the semi-infinite media are
$T_1=400$ K and $T_2=300$ K and the width of the vacuum gap is $10$ $\mu$m.

Figure \ref{fig:temp} shows the LDOS for the electric, magnetic, and total EM fields
and the corresponding effective field temperatures as a function of position.
The electric LDOS in Fig.~\ref{fig:temp}(a) oscillates in the vacuum and saturates to
constant values in the lossy media far from the interfaces reflecting the formation of
partial standing waves due to interference. The oscillation
of the electric LDOS inside the cavity is strongest at resonant energies $\hbar\omega=0.056$ eV ($\lambda=22.1$ $\mu$m),
$\hbar\omega=0.118$ eV ($\lambda=10.5$ $\mu$m), and $\hbar\omega=0.180$ eV ($\lambda=6.89$ $\mu$m).
The oscillations in the electric LDOS manifest
the Purcell effect and the related position-dependent strength of the field-matter coupling
of particles placed in the cavity.
The magnetic LDOS in Fig.~\ref{fig:temp}(b) also oscillates inside the cavity.
However, the positions of the peaks coincide with the minima of the electric LDOS in
Fig.~\ref{fig:temp}(a). In contrast to electric LDOS, the magnetic LDOS reaches its maximum
values within the semi-infinite media due to the low finesse cavity and
different dependence on the refractive index
as can be seen in the latter form in Eq.~\eqref{eq:bldos} compared to the
electric LDOS in Eq.~\eqref{eq:eldos}. The total EM LDOS in Fig.~\ref{fig:temp}(c)
is constant with respect to position inside the cavity. This illustrates how the
EM energy oscillates between its electric and magnetic forms in a way preserving
the total energy.
However, the total LDOS is position-dependent and oscillatory near interfaces
inside the lossy media since the electric and magnetic LDOSs are not equal due to
the bound states related to the material polarizability.

The effective field temperature defined using Eq.~\eqref{eq:temperature} is plotted
for the electric field in Fig.~\ref{fig:temp}(d). It has a strong position dependence
and it oscillates both in the vacuum and inside the lossy media. The position dependence originates
from the nonuniform coupling to the two thermal reservoirs. In the lossy media
the oscillations are damped and the effective electric field temperature saturates to
the reservoir temperature far from the interfaces. The characteristic distance for the damping
of the oscillations depends on the
photon energy and the material absorptivity and, in the selected example structure,
it has a typical value of the order of 10 $\mu$m
increasing for smaller photon energies and decreasing for larger photon energies.
Since the field-matter interaction takes place through the electric field, the local electric
field temperature directly reveals the local material temperature required for having
no net heat energy exchange between the field and matter making it an experimentally
measurable quantity as discussed in Sec.~\ref{sec:balance}.
The magnetic field temperature is plotted in Fig.~\ref{fig:temp}(e). It also has a strong
position dependence, but the peaks are located at different positions when compared to the
electric field temperature in Fig.~\ref{fig:temp}(d).

\begin{figure*}
\includegraphics[width=\textwidth]{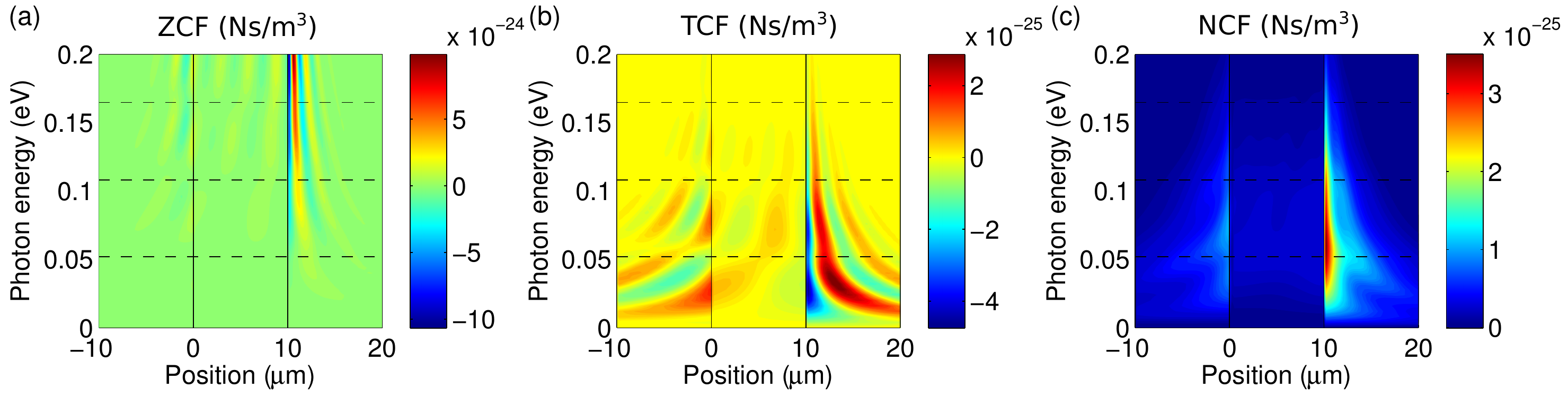}
\caption{\label{fig:force}(Color online) (a) The zero-point, (b) thermal, and (c) nonequilibrium
Casimir force contributions of the
spectral force density in the case of a 10-$\mu$m wide lossy cavity with refractive index $n_\mathrm{c}=1.1+0.1i$.
The left and right cavity boundaries have refractive indices
$n_1=1.5+0.3i$ and $n_2=2.5+0.5i$ and temperatures 400 and 300 K.
Solid lines denote the boundaries of the cavity and dashed lines denote resonant energies.
}
\end{figure*}

The total EM
field temperature in Fig.~\ref{fig:temp}(f) is constant with respect to position inside
the cavity as the total EM LDOS in Fig.~\ref{fig:temp}(c). The position independence
of the total field temperature follows from the position independence of the total EM
photon number in the vacuum as discussed in Sec.~\ref{sec:theory}. In contrast to the electric
and magnetic field temperatures in Figs.~\ref{fig:temp}(d) and \ref{fig:temp}(e),
the changes of the total EM field temperature and photon number near interfaces are allways
monotonic with respect to position, which is an expected result
for the photon number of the total EM field.

\vspace{-0.4cm}
\subsection{Electromagnetic forces}
\vspace{-0.4cm}

Next we study the Casimir force densities in a lossy cavity filled with
material having refractive index $n_\mathrm{c}=1.1+0.1i$. As before, the cavity width
is 10 $\mu$m and the left and right cavity boundaries have
refractive indices $n_1=1.5+0.3i$ and $n_2=2.5+0.5i$ and temperatures 400 and 300 K.
In contrast to the vacuum cavity case, the lossy cavity medium
acts as an additional field source emitting photons. In this work, we
calculate the temperature of the lossy cavity medium self-consistently
so that the photon emission equals absorption at every point, i.e., the net
emission rate in Eq.~\eqref{eq:divP} is zero. This also means that other
heat conduction mechanisms than radiation are neglected.

The force density contributions calculated by using Eq.~\eqref{eq:forcedensity2}
are plotted in Fig.~\ref{fig:force} as a function
of the position and photon energy.
Since the refractive index forms step functions at the interfaces, it follows from definitions
\eqref{eq:totaln} and \eqref{eq:uldos} that the total EM LDOS and the total
EM photon number are also discontinuous so that
the force density contributions contain delta functions at the interfaces. This can not be seen
in the figures and it will give an additional contribution when integrating the total force.
The spectral component of the
zero-point Casimir force is shown in Fig.~\ref{fig:force}(a), the
thermal Casimir force in Fig.~\ref{fig:force}(b), and the nonequilibrium
Casimir force in Fig.~\ref{fig:force}(c).
The zero-point Casimir force dominates especially in the case of high frequencies
except close to its zeros where the thermal and nonequilibrium contributions
become important. The thermal and nonequilibrium contributions
decay at high frequencies due to the smaller thermal excitation of the high energy states.
In contrast, the zero-point Casimir force does not decay but increases with
frequency since the materials are assumed to be nondispersive in the studied frequency range.
One can also see that the zero-point and thermal Casimir contributions both obtain
positive and negative values corresponding to forces to the left and right depending on the
position and photon energy. In contrast, the nonequilibrium Casimir force is always positive
since the derivative of the photon number is negative in Eq.~\eqref{eq:forcedensity2} because
the total photon number monotonically decreases towards the colder medium.
The direction and magnitude of the total force density depends on the
value of the integral when the spectral force density is integrated over the frequency axis.
The observable total force exerted on a volume is the sum of the three force density
components integrated over the volume.
\vspace{0.3cm}

\begin{figure*}
\includegraphics[width=\textwidth]{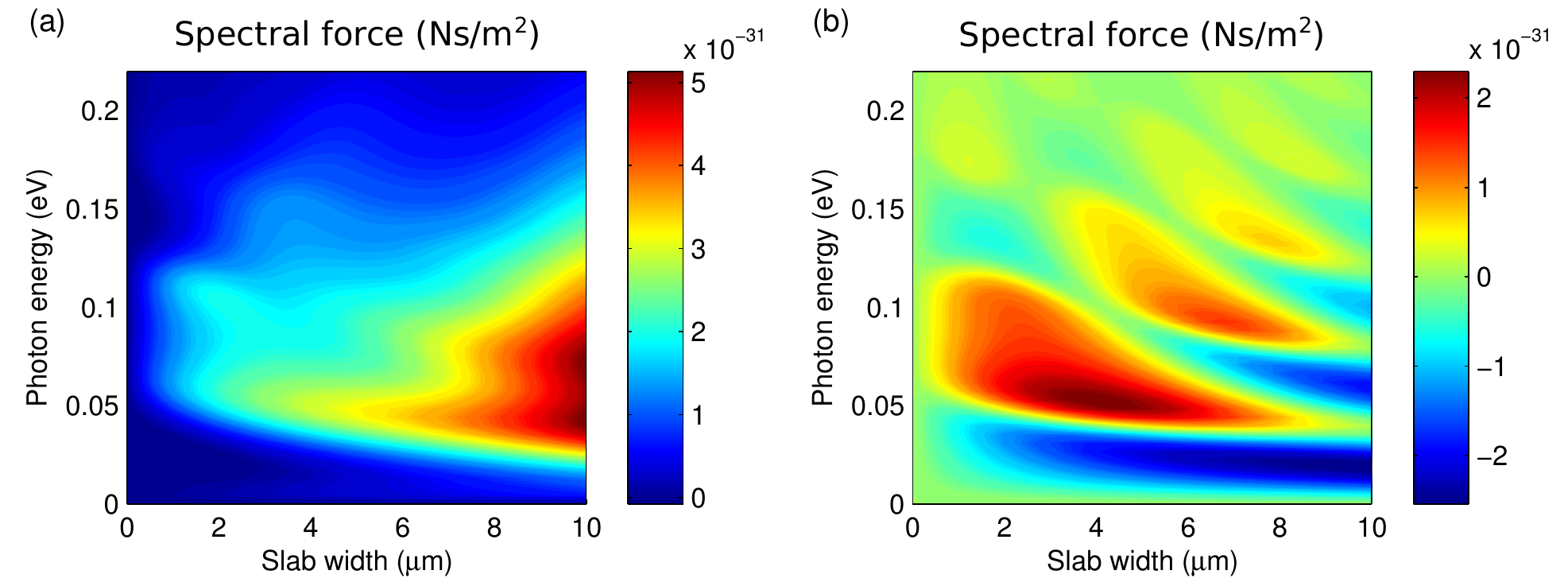}
\caption{\label{fig:pres}(Color online) (a) The spectral force exerted on a lossy material slab
with refractive index $n_\mathrm{slab}=1.5+0.3i$ inside the 10-$\mu$m wide vacuum cavity. (b) The
spectral force in the case of a lossless material slab with refractive index $n_\mathrm{slab}=1.5$.
The left and right cavity boundaries have equal refractive indices $n_1=n_2=2.5+0.5i$ and
temperatures 400 and 300 K. The center of the slab is positioned in the middle of the cavity.
}
\end{figure*}

To learn more on the EM pressure under conditions where the well-known zero-point Casimir force vanishes
we study the force exerted on lossy and lossless material slabs placed in the middle of
a symmetric vacuum cavity. In the case of a slab in the middle of a symmetric cavity,
the force is nonzero only for setups that are not under thermal equilibrium
since the zero-point Casimir contribution to the force cancels resulting in the force in Eq.~\eqref{eq:forceslab2}.
In contrast to the previous geometry, the refractive indices of cavity boundaries
are equal $n_1=n_2=2.5+0.5i$, but the left and right reservoir temperatures are 400 and 300 K as before.

The total spectral force experienced by the lossy slab with refractive index $n_\mathrm{slab}=1.5+0.3i$
is plotted in Fig.~\ref{fig:pres}(a) as a function of the slab width and photon energy.
The force is always directed towards the colder material.
In the limit of zero slab width, the force goes to zero since the volume of the interacting
slab material disappears. When the slab width is increased, also the force increases.
However, the force disappears for low and high frequencies as the photon energy $\hbar\omega$
or the photon probability given by the Bose-Einstein distribution becomes zero. Also note
that the force becomes zero for all slab widths and photon energies
if the left and right reservoir temperatures are equal.

The total force experienced by the lossless slab with real refractive index $n_\mathrm{slab}=1.5$
is plotted in Fig.~\ref{fig:pres}(b). Comparison with the case of a lossy slab in Fig.~\ref{fig:pres}(a)
shows that the force strongly depends on losses.
The force on a lossless slab can clearly obtain negative values for some frequencies and
slab widths indicating that the force is unexpectedly towards the medium at higher temperature
while the energy flow according to the Poynting vector is towards the medium at lower temperature.
This is called optical pulling force \cite{Chen2011} and it is expected to be possible because
of the combined effect of cavity resonances and the
proportionally larger contribution of the low energy photons of the lower temperature thermal
reservoir.
\vspace{0.2cm}

\section{\label{sec:conclusions}Conclusions}
\vspace{-0.4cm}

The quantization approach we recently introduced to define position-dependent
ladder operators to describe the quantized electric field and light-matter interactions
was extended to also describe the quantization of the magnetic and total EM fields. 
The quantization is based on defining electric, magnetic, and total EM photon ladder operators that
by definition obey the canonical commutation relations.
The previous electric-field-based quantization was only able to describe
the electric field photon number and local
energy balance, whereas
the generalized approach also allows consistent
quantum optical description of 
the magnetic and total EM field photon numbers, energy balance of magnetic interactions, radiation pressure,
and EM forces in lossy and dispersive structures.
One additional strength of the formalism is that
formulas involving photon-number expectation values are expected to generalize to fields
with any kind of quantum statistics, such as single photon and laser fields.

We have studied the energy flow and induced EM forces in cavity structures with cavity walls at different temperatures. 
Our results show that the electric and magnetic field operators and temperatures are generally position dependent under
nonequilibrium conditions due to being defined in terms of the position-dependent noise
fields generating them. However, the photon number associated with the total EM field is
always constant in lossless media as one might also intuitively expect from a quantum number
describing the energy density of the total EM field.
Furthermore, the direction of the EM force in the cavity was shown to be position dependent producing 
optical pulling and pushing forces at different cavity positions.

We expect that our model will give insight in studying the optical energy transfer in 
nanodevices as well as modeling optomechanical devices. In addition, extending the formalism to
quantum systems that are not limited to thermal fields could possibly enable a versatile
approach to model various quantum optical experiments.

\vspace{-0.3cm}
\begin{acknowledgments}
\vspace{-0.4cm}
This work has in part been funded by the Academy of Finland and the Aalto Energy Efficiency Research Programme.
\end{acknowledgments}

\bibliography{bibliography}

\end{document}